\documentclass{aa}

\usepackage{natbib}
\usepackage{longtable}
\usepackage{lscape}
\usepackage{amsmath}
\usepackage{txfonts}
\usepackage{color}
\usepackage{url}
\usepackage{xspace}
\usepackage{rotating}

\usepackage{graphicx}
\bibpunct{(}{)}{;}{a}{}{,}
\newcommand{\corot}{\textsl{CoRoT}\xspace}

\begin{document}

  \title{Planetary transit candidates in \corot-IRa01 field\thanks{The \corot space mission, launched on December 27th 2006, 
  has been developed and is operated by CNES, with contributions from Austria, Belgium, Brazil, ESA,
  Germany, and Spain. Four 
  French laboratories associated with the CNRS (LESIA, LAM, IAS ,OMP) collaborate with CNES on the satellite development.
  First CoRoT data are available to the public from the CoRoT archive: http://idoc-corot.ias.u-psud.fr.}    }
  
  \author{S.~Carpano \inst{1}
    \and J.~Cabrera \inst{2,3}
    \and R.~Alonso\inst{4}
    \and P.~Barge\inst{4}
\and S.~Aigrain \inst{6}
\and J.-M.~Almenara \inst{7}
\and P.~Bord\'e \inst{8}
\and F.~Bouchy \inst{9}
\and L.~Carone \inst{10}
\and H.~J.~Deeg \inst{7}
\and R.~De la Reza  \inst{11}
\and M.~Deleuil \inst{4}
\and R.~Dvorak \inst{12}
\and A.~Erikson \inst{2}
\and F.~Fressin \inst{22}
\and M.~Fridlund  \inst{1}
\and P.~Gondoin \inst{1}
\and T.~Guillot  \inst{13}
\and A.~Hatzes \inst{14}
\and L.~Jorda  \inst{4}
\and H.~Lammer  \inst{15}
\and A.~L\'eger  \inst{8}
\and A.~Llebaria \inst{4}
\and P.~Magain \inst{16}
\and C.~Moutou  \inst{4}
\and A.~Ofir  \inst{20}
\and M.~Ollivier  \inst{8}
\and E.~J.~Pacheco \inst{21}
\and M.~P\"atzold  \inst{10}
\and F.~Pont \inst{6}
\and D.~Queloz  \inst{5}
\and H.~Rauer  \inst{2}
\and C.~R\'egulo \inst{7}
\and S.~Renner \inst{2,17,18}
\and D.~Rouan  \inst{19}
\and B.~Samuel \inst{8}
\and J.~Schneider  \inst{3}
\and G.~Wuchterl \inst{14}
}

  \offprints{S. Carpano, e-mail: scarpano@rssd.esa.int}
  \institute{Research and Scientific Support Department, ESTEC/ESA, PO Box 299, 2200 AG Noordwijk, The Netherlands
    \and
    Institute of Planetary Research, German Aerospace Center, Rutherfordstrasse 2, 12489 Berlin, Germany 
    \and 
    LUTH, Observatoire de Paris, CNRS, Universit\'e Paris Diderot; 5 place Jules Janssen, 92190 Meudon, France 
    \and
    Laboratoire d'\,Astrophysique de Marseille, UMR 6110,  38 rue F. Joliot-Curie, 13388 Marseille, France
\and
Observatoire de Gen\`eve, Universit\'e de Gen\`eve, 51 chemin des Maillettes, 1290 Sauverny, Switzerland
\and
School of Physics, University of Exeter, Stocker Road, Exeter EX4 4QL, United Kingdom
\and
Instituto de Astrof\'isica de Canarias, E-38205 La Laguna, Tenerife, Spain
\and
Institut d' Astrophysique Spatiale, Universit\'e Paris XI, F-91405 Orsay, France
\and    
Institut d'Astrophysique de Paris, Universit\'e Pierre \& Marie Curie, 98bis Bd Arago, 75014 Paris, France
\and
Rheinisches Institut f\"ur Umweltforschung an der Universit\"at  zu K\"oln, Aachener Strasse 209, 50931, Germany
\and
Observat\'orio Nacional, Rio de Janeiro, RJ, Brazil
\and
University of Vienna, Institute of Astronomy, T\"urkenschanzstr. 17, A-1180 Vienna, Austria
\and
Observatoire de la C\^ote d'\,Azur, Laboratoire Cassiop\'ee, BP 4229, 06304 Nice Cedex 4, France
\and
Th{\"u}ringer Landessternwarte, Sternwarte 5, Tautenburg 5, D-07778 Tautenburg, Germany
\and
Space Research Institute, Austrian Academy of Science, Schmiedlstr. 6, A-8042 Graz, Austria
\and
University of Li\`ege, All\'ee du 6 ao\^ut 17, Sart Tilman, Li\`ege 1, Belgium
\and
Laboratoire d'Astronomie de Lille, Universit\'e de Lille 1, 1 impasse de l'Observatoire, 59000 Lille, France
\and
Institut de M\'ecanique C\'eleste et de Calcul des Eph\'em\'erides, UMR 8028 du CNRS, 77 avenue Denfert-Rochereau, 75014 Paris, France
\and
LESIA, Observatoire de Paris-Meudon, 5 place Jules Janssen, 92195 Meudon, France
\and 
School of Physics and Astronomy, Raymond and Beverly Sackler Faculty of Exact Sciences, Tel Aviv University, Tel Aviv, Israel
\and 
Instituto de Astronomia, Geof\'isica e Ci\^encias Atmosf\'ericas, Universidade de S$\tilde{\textrm{a}}$o Paulo, 05508-900
S$\tilde{\textrm{a}}$o Paulo, Brazil
\and
Harvard University, Department of Astronomy, 60 Garden St., MS-16, Cambridge, MA 02138, USA
} 
  
  \date{Submitted: XXX; Accepted: XXX}
 
  \abstract
      {\corot is a pioneering space mission devoted to the analysis
	of stellar variability and the photometric detection of
	extrasolar planets.}
      {We present the list of planetary transit candidates 
	detected in the first field observed by CoRoT, IRa01, the
	initial run toward the Galactic anticenter, which lasted
	for 60\,days.} 
      {We analysed 3898 sources in the coloured bands and 5974 in the
	monochromatic band. Instrumental noise and 
	stellar variability were taken into account using  detrending tools 
	before applying various transit search algorithms.}
      {Fifty sources were classified as planetary
	transit candidates and the most reliable 40 detections were declared targets for follow-up ground-based observations. 
	Two of these targets have so far been confirmed as planets, \corot-1b and \corot-4b,
	for which a complete characterization and specific studies were performed.}  
      {}
     
  \maketitle

%
%____________________________________________________________________________
\section{Introduction}
\label{sec:intro}

The transit method for detecting exoplanets identifies candidates by monitoring stars 
for long periods of time, then processing the data to isolate stars that exhibit a periodic 
flux drop consistent with a Jupiter-sized or smaller companion passing between its parent star 
and the observer. A large number of targets is necessary, because the probability of a planet 
producing an observable transit is very low, due to geometric effects. The processing and 
analysis of gathered data is thus a major undertaking.

\begin{figure}
 \resizebox{\hsize}{!}{\includegraphics[angle=90]{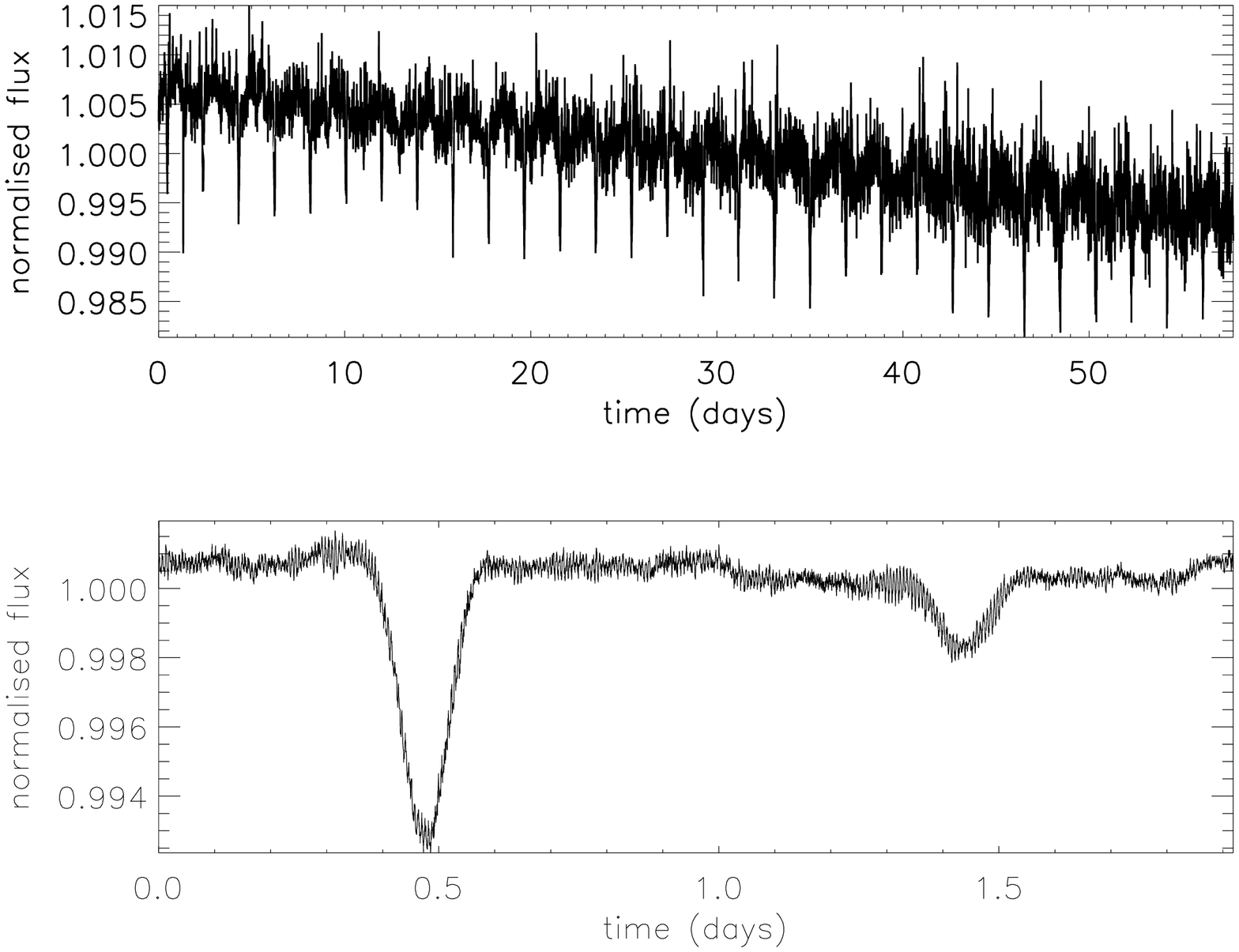}} \caption{An eclipsing binary found in IRa01 showing small secondary eclipses. 
 Raw (top), smoothed, and detrended folded light curve (bottom).} 
\label{fig:binary1}
 \resizebox{\hsize}{!}{\includegraphics[angle=90]{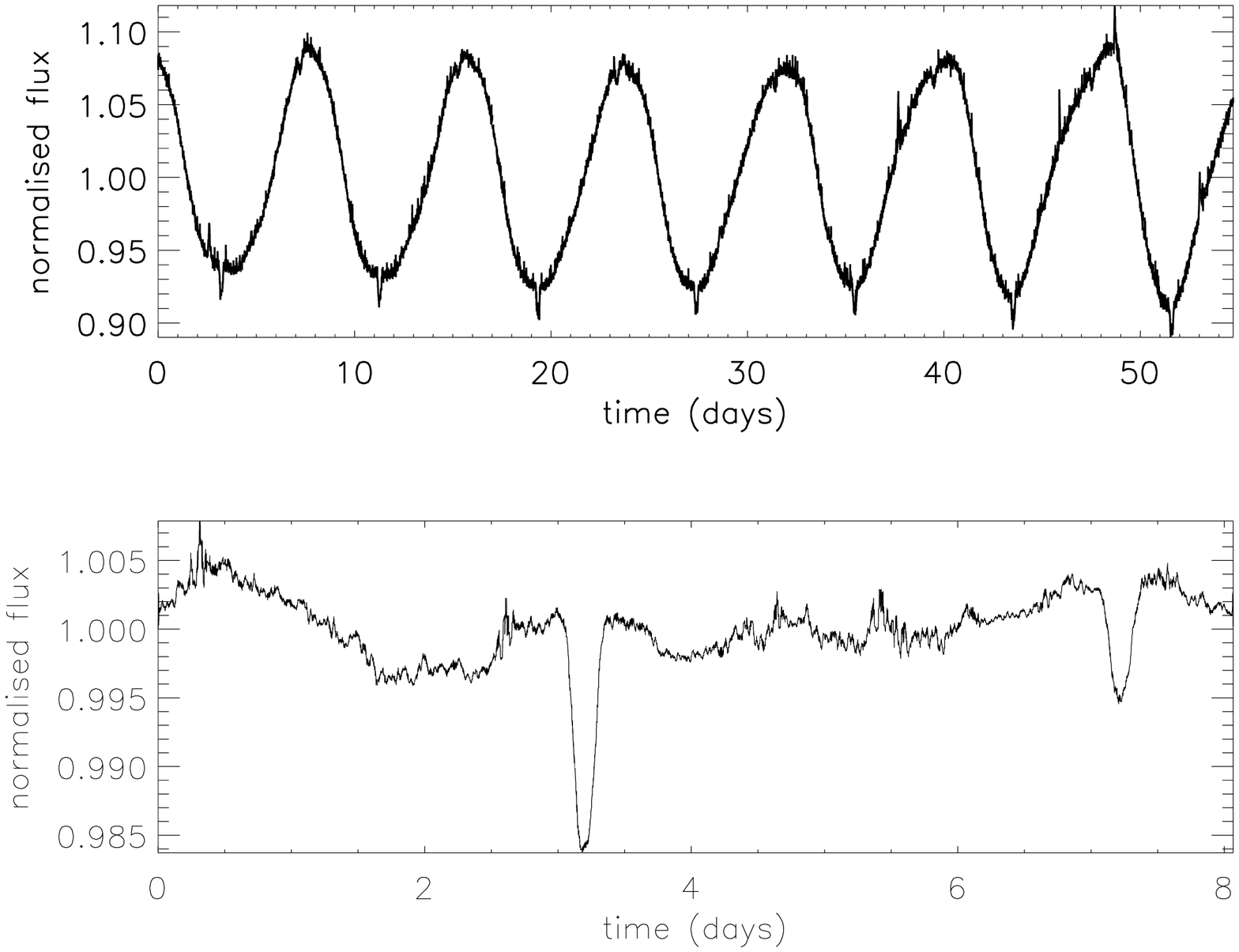}} \caption{An eclipsing binary found in IRa01 showing in phase modulation.
 Raw (top), smoothed, and detrended folded light curve (bottom).} 
\label{fig:binary2}
 \resizebox{\hsize}{!}{\includegraphics[angle=90]{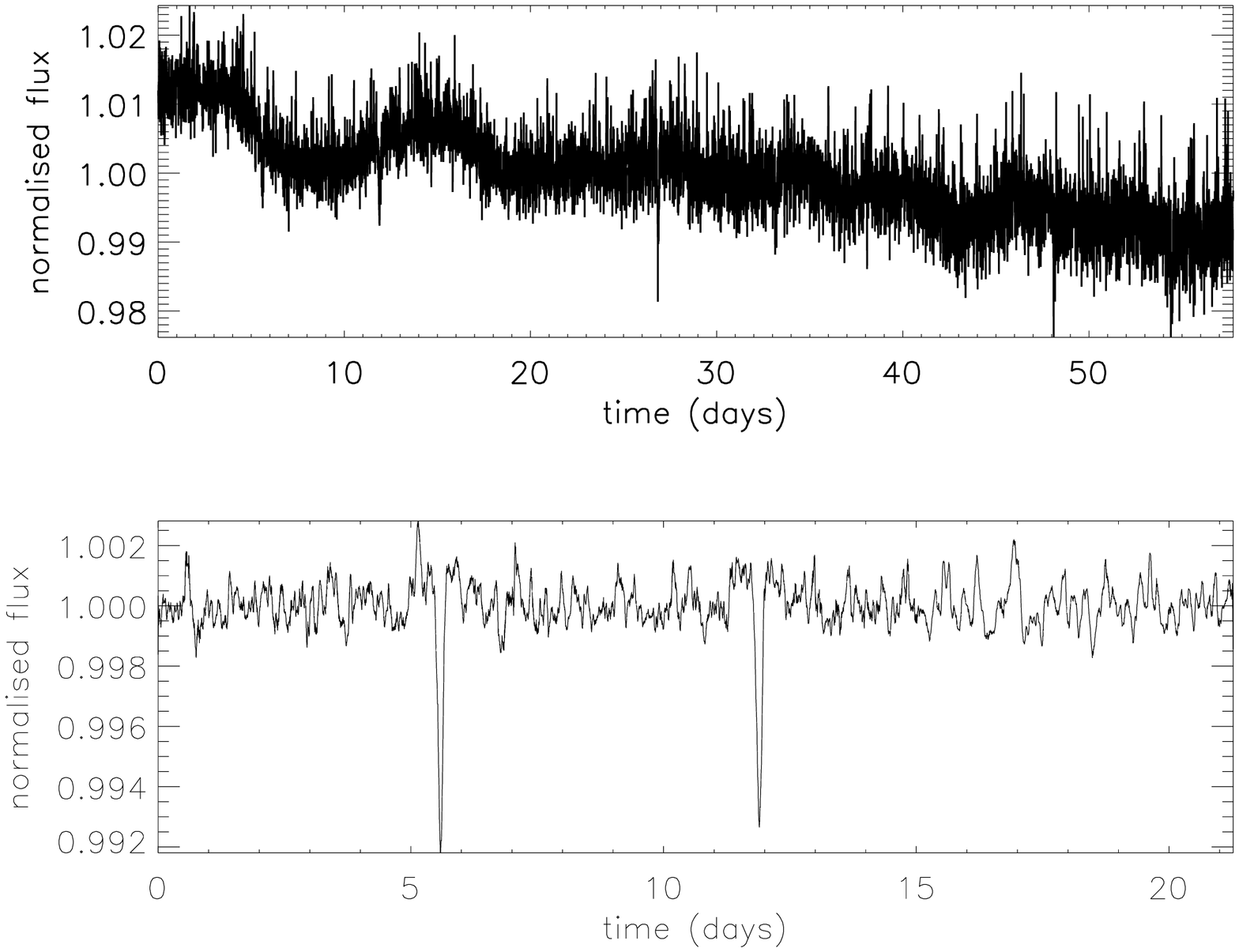}} \caption{An eclipsing binary found in IRa01 showing orbital eccentricities.
 Raw (top), smoothed, and detrended folded light curve (bottom).} 
\label{fig:binary3}
\end{figure}

\begin{figure}
 \resizebox{\hsize}{!}{\includegraphics[angle=90]{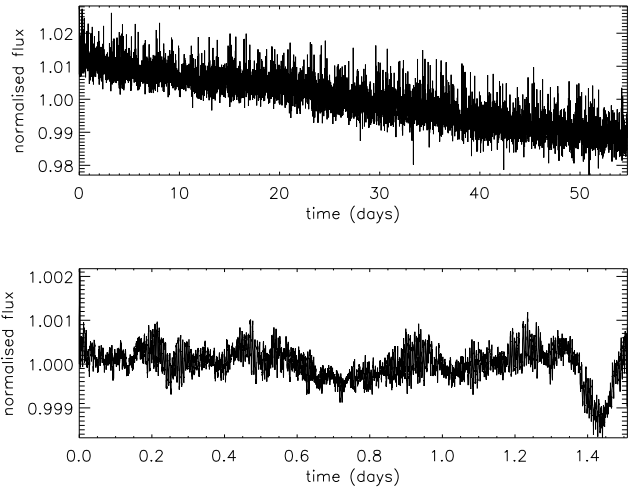}} \caption{Raw and folded light curve of a planetary candidate (source No. 46, E2 4124).} 
\label{fig:good}
\end{figure}

The methodology used to analyse
thousands of light curves in the search for transiting extrasolar
planets was described in detail by \citet{Gould2006} for OGLE
data. We summarize here a few concepts:
\begin{itemize}
\item \corot light curves are processed and filtered for instrumental
  noise as described in \citet{Drummond2008};
\item each of the detection teams applies its own algorithms for
 detrending the signal (e.g., variability, noise) and searching for
 planetary transits (see \citealt{Moutou2005,Moutou2007});
\item the results of each team are combined and each candidate is discussed 
  individually.
\end{itemize}
In our final discussion, a check is performed to reject clear
eclipsing binaries, i.e, systems with lights curves that exhibit secondary eclipses, 
out-of-transit photometric modulations, and/or events that are too deep to be caused 
by transiting planets.  The shape of transits is also
analysed: photometric dips of planets  have a ``U" shape, while
binaries are more ``V" shaped. These criteria, however, can only be used
for data of with relatively high signal-to-noise ratios. Some examples of 
eclipsing binary light curves are shown in Figs.~\ref{fig:binary1}, \ref{fig:binary2} and \ref{fig:binary3}. 
Raw light curves are shown in the top panel and smoothed, detrended, and folded
light curves are shown in the bottom panel.
These exhibit the typical features of small secondary eclipses, in phase modulation, 
and secondary transits out of phase 0.5.
Figure~\ref{fig:good} shows the raw and folded light curves of a good planetary candidate with a shallow transit 
(source No. 46, E2 4124, in Tables~\ref{tab:cand} and \ref{tab:param}).

Source confusion with background binaries will also produce false
candidates; this is true in particular for \corot because of its large PSF \citep{Barge2008,Drummond2008}. 
In this case, one benefits from the three coloured bands of the CoRoT photometric mask. 
When a candidate is bright enough for its flux to be separated into three 
bands/colours, the transit is occasionally not observed in one or more of the bands/colours 
or is a significantly different depth in the separate bands/colours.
For the remaining candidates, photometric and/or spectroscopic follow-up are essential to 
determining of the masses of the system components, by measurements of the radial velocity 
shift of the spectral lines of the parent star that occur as the planet orbits.
In the case of \corot, photometric follow-up is useful in cases of source
confusion. Spectroscopic ground-based measurements, on the other hand
are essential for determinating the masses of the system components, 
via measurements of the radial velocity shift of the spectral lines of the parent star 
that occur as the planet orbits.

In this work, we present the results of the joint work of the \corot detection 
teams, a huge effort to separate the wheat from the chaff  to provide
accurate parameters for the interesting objects. 
The IRa01 \corot data are now public. We offer the fruits of our labor to the 
astronomical community so it may serve as a starting point for interested researchers.
Section~\ref{sec:data} contains some details about this initial CoRoT run, including the candidate 
information from the satellite itself. In Sect.~\ref{sec:trparam},
we provide the list of the 50 transiting candidates observed in the 
first \corot field IRa01 and their transit parameters. 
Results are summarized in Sect.~\ref{sec:conc}. 

%_____________________________________________________________________________

\section{\corot observations of IRa01 field}
\label{sec:data}
\begin{figure}
 \resizebox{\hsize}{!}{\includegraphics[bb= 37 192 578
 627,clip=true]{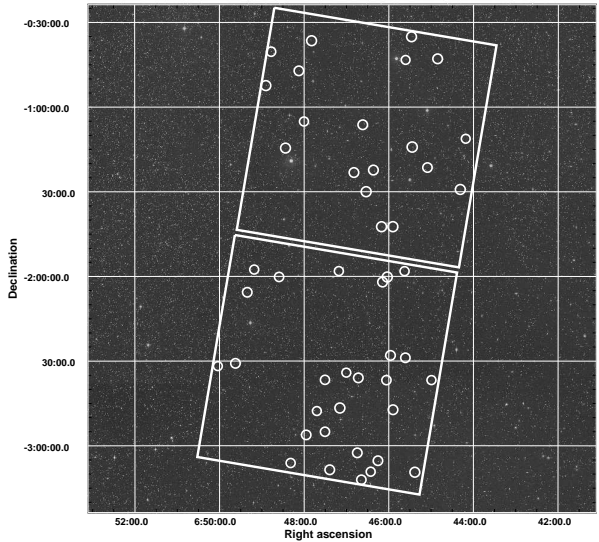}} \caption{DSS image of the sky observed by
 \corot during the IRa01. Overlaid are the positions of the 50
 planetary transit candidates and the portion of the field covered by
 the 2 exoplanets CCD.} 
\label{fig:image}
\end{figure}

\corot observed its first field from early February 2008 until early
April, for approximatively 60\,days. The run code `IRa01' is explained
as following. The `IR' means `initial run' in contrast to the subsequent
`long runs' (LR) and `short runs' (SR). The third letter refers to the
direction with respect to the Galactic center (`a', as in this case,
anticenter or `c' Galactic center). The last two digits are the sequence 
for this type of observation (01 being the first one). 

\begin{table}
 \centering
 \caption{List of the detection teams (institutes and people).}
  \label{tab:teams}
  \begin{tabular}{cl}
  \hline
    team     & participants \\
    \hline
    DLR      & Heike Rauer, Anders Erikson, Stefan Renner \\
    ESTEC    & Malcolm Fridlund, Stefania Carpano \\
    Exeter   & Suzanne Aigrain, Fr{\'e}d{\'e}ric Pont, Aude Alapini \\
    IAC      & Hans Deeg, Jos{\'e} M. Almenara, Clara R\'egulo \\
    IAS      & Pascal Bord{\'e}, Benjamin Samuel \\
    K{\"o}ln & Martin P{\"a}tzold, Ludmilla Carone \\
    LAM      & Pierre Barge, Roi Alonso \\
    LUTh     & Jean Schneider, Juan Cabrera \\
    \hline
  \end{tabular}
\end{table}

3898 sources were observed in IRa01 using coloured filters (B, V,
R colours), while 5974 sources were monitored at a single
monochromatic band. To analyse these data sets, detection 
teams were established in a number of different collaborating institutes.
Their task is to provide a list of candidates, their ranking 
(according to the probability of their planetary nature), 
as well as a first estimate of transit ephemerides and parameters.
At this point, 8 teams are participating, each using their own independently-developed 
detection methods.
Table~\ref{tab:teams} contains a list of the different institutions involved and 
the names of the contributors.
Some of these methods were presented during a pre-launch performance simulation
described in \cite{Moutou2005}, while others have been or will be developed in separated papers
\citep[i.e.,][]{Carpano2008, Renner2008, Regulo2007}.  
The algorithms described in these works are generally based on the following 
fundamental approaches: correlation with sliding transit template, box-shaped signal search, 
box-fitting least-squares (BLS),
wavelet transformation, or Gaussian fitting of folded light curve.                                 
A merged list of 92 planetary transit candidates was compiled by
the teams, 
which was reduced to a final list of 50 candidates after discussion 
(most of the other 42 candidates were classified as binaries). The 40 most robust candidates 
were recommended for ground-based follow-up, the results of which are reported in 
Moutou~et~al.~2009 (accepted).
Two planets from the final list of 40 candidates, \corot-1b and \corot-4b, have so far been confirmed as planets.
More details about the discovery of these two planets can be found in \cite{Barge2008b} and \cite{Aigrain2008}, respectively.

Figure~\ref{fig:image} shows the sky coverage of the 2 CCDs dedicated to 
exoplanetary science and the positions of the candidates within this field of view.
Table~\ref{tab:cand} provides a list of planetary candidates in the IRa01 field, 
including their CoRoT- and window-ID numbers, J2000 positions, an indication of  whether the candidate was observed 
in three colours (``CHR") or monochrome (``MON"), magnitude(s), and exposure times (in s). 
A change in the time sampling from 512\,s to 32\,s
indicates that several transits were detected in the first portion of
the light curve and the Alarm Mode \citep{Quentin2006, Surace2008} was chosen to resample those targets      
to improve the time accuracy. All parameters derived
from the Exo-Dat database \citep{Meunier2007, Deleuil2008} .

\begin{table*}
 \centering
 \caption{List of the 50 planetary transit candidates detected in the \corot IRa01 field, see text for more details.} 
  \label{tab:cand}
 \begin{tabular}{rrrrrrrr}
 \hline
  No. &\textsl{CoRoT}-ID & Win-ID & right ascension, declination & Exo-Dat R mag & Colour & Time Sampling\\
 \hline  
 1 & 0102723949 &  E1 2046  & 6:44:11.03943,-1:11:13.236  &   13.60  & CHR  & 512     \\
 2 & 0102729260 &  E1 1319  & 6:44:18.72070,-1:29:11.328  &   14.77  & CHR  & 512     \\
 3 & 0102763847 &  E1 1158  & 6:45: 5.28076,-1:21:25.236  &   13.11  & CHR  & 512, 32 \\
 4 & 0102787048 &  E1 0288  & 6:45:36.24023,-0:43:17.400  &   13.17  & CHR  & 512, 32 \\
 5 & 0102787204 &  E2 3787  & 6:45:36.48010,-2:28:50.664  &   14.00  & CHR  & 512     \\
 6 & 0102798247 &  E2 1857  & 6:45:53.99963,-2:47:15.900  &   14.06  & CHR  & 512, 32 \\
 7 & 0102806520 &  E1 4591  & 6:46:10.31982,-1:42:23.688  &   13.70  & CHR  & 512     \\
 8 & 0102809071 &  E2 1136  & 6:46:15.36072,-3: 5:19.608  &   13.20  & CHR  & 512, 32 \\
 9 & 0102815260 &  E2 2430  & 6:46:25.68054,-3: 9:13.284  &   14.57  & CHR  & 512, 32 \\
10 & 0102825481 &  E2 0203  & 6:46:43.20007,-2:35:58.308  &   13.07  & CHR  & 512, 32 \\
11 & 0102826302 &  E2 1712  & 6:46:44.63928,-3: 2:32.208  &   13.98  & CHR  & 512, 32 \\
12 & 0102829121 &  E1 0399  & 6:46:49.44031,-1:23:11.616  &   13.66  & CHR  & 512     \\
13 & 0102855534 &  E2 1736  & 6:47:30.47974,-2:55: 4.116  &   13.85  & CHR  & 512, 32 \\
14 & 0102856307 &  E1 0396  & 6:47:31.68091,-1:23:26.808  &   13.58  & CHR  & 512     \\
15 & 0102874481 &  E2 1677  & 6:47:57.11975,-2:56:10.896  &   13.84  & CHR  & 512     \\
16 & 0102890318 &  E2 1126  & 6:48:19.20044,-3: 6: 7.776  &   13.43  & CHR  & 512, 32 \\
17 & 0102895957 &  E1 0783  & 6:48:26.40015,-1:14:31.344  &   12.72  & CHR  & 512, 32 \\
18 & 0102912369 &  E1 0330  & 6:48:46.79993,-0:40:21.972  &   13.45  & CHR  & 512, 32 \\
19 & 0102918586 &  E1 2755  & 6:48:54.23950,-0:52:22.800  &   12.24  & CHR  & 512, 32 \\
20 & 0102753331 &  E1 4617  & 6:44:50.87952,-0:42:53.280  &   15.87  & MON  & 512     \\
21 & 0102759638 &  E2 3724  & 6:44:59.52026,-2:36:45.144  &   14.79  & MON  & 512     \\
22 & 0102777119 &  E2 4290  & 6:45:23.04016,-3: 9:23.688  &   15.03  & MON  & 512     \\
23 & 0102779966 &  E1 4108  & 6:45:26.87988,-1:14: 9.456  &   14.94  & MON  & 512     \\
24 & 0102780627 &  E1 1531  & 6:45:27.83936,-0:35: 4.668  &   14.98  & MON  & 512     \\
25 & 0102788073 &  E2 2009  & 6:45:37.67944,-1:58: 9.300  &   14.18  & MON  & 512     \\
26 & 0102798429 &  E1 2774  & 6:45:54.23950,-1:42:22.752  &   15.52  & MON  & 512     \\
27 & 0102800106 &  E2 3010  & 6:45:57.59949,-2:28: 0.732  &   15.49  & MON  & 512     \\
28 & 0102802430 &  E2 4300  & 6:46: 2.16064,-2: 0:13.428  &   14.33  & MON  & 512     \\
29 & 0102802996 &  E2 3150  & 6:46: 3.35999,-2:36:46.548  &   14.96  & MON  & 512     \\
30 & 0102805893 &  E2 2604  & 6:46: 8.88062,-2: 2: 0.348  &   15.42  & MON  & 512     \\
31 & 0102812861 &  E1 2648  & 6:46:21.84082,-1:22:19.128  &   15.59  & MON  & 512     \\
32 & 0102819021 &  E1 2328  & 6:46:32.16064,-1:29:58.812  &   15.10  & MON  & 512     \\
33 & 0102821773 &  E1 4998  & 6:46:36.95984,-1: 6:15.768  &   15.48  & MON  & 512     \\
34 & 0102822869 &  E2 4058  & 6:46:38.88062,-3:12: 4.860  &   15.55  & MON  & 512     \\
35 & 0102835817 &  E2 3425  & 6:47: 0.23987,-2:34: 7.140  &   15.60  & MON  & 512     \\
36 & 0102841669 &  E2 3854  & 6:47: 9.36035,-2:46:39.108  &   14.91  & MON  & 512     \\
37 & 0102842120 &  E2 3952  & 6:47:10.07996,-2:57: 1.944  &   13.98  & MON  & 512     \\
38 & 0102842459 &  E2 1407  & 6:47:10.79956,-1:58: 7.356  &   14.77  & MON  & 512     \\
39 & 0102850921 &  E2 2721  & 6:47:23.75977,-3: 8:32.424  &   12.90  & MON  & 512     \\
40 & 0102855472 &  E2 0704  & 6:47:30.47974,-2:36:40.140  &   13.86  & MON  & 512     \\
41 & 0102863810 &  E2 4073  & 6:47:42.00073,-2:47:43.404  &   15.39  & MON  & 512     \\
42 & 0102869286 &  E1 2329  & 6:47:49.44031,-0:36:29.052  &   15.52  & MON  & 512     \\
43 & 0102876631 &  E1 3336  & 6:48: 0.23987,-1: 5: 5.964  &   14.56  & MON  & 512     \\
44 & 0102881832 &  E1 4911  & 6:48: 7.43958,-0:47: 9.024  &   15.05  & MON  & 512     \\
45 & 0102903238 &  E2 4339  & 6:48:35.52063,-2: 0:12.096  &   15.61  & MON  & 512     \\
46 & 0102926194 &  E2 4124  & 6:49: 3.59985,-2:48:34.488  &   15.73  & MON  & 512     \\
47 & 0102932089 &  E2 3819  & 6:49:10.79956,-1:57:34.452  &   16.00  & MON  & 512     \\
48 & 0102940315 &  E2 4467  & 6:49:20.63965,-2: 5:37.716  &   15.81  & MON  & 512     \\
49 & 0102954464 &  E2 3856  & 6:49:37.43958,-2:30:49.140  &   15.97  & MON  & 512     \\
50 & 0102973379 &  E2 1063  & 6:50: 2.40051,-2:31:47.604  &   14.09  & MON  & 512     \\

\hline
  \end{tabular}
\end{table*}

\section{Compiling a list of candidates with their transit parameters}
\label{sec:trparam}    
The selection process of planetary candidates for follow-up has several steps. First, 
each detection team analyses the tens of thousands of light curves independently using their 
own filtering and detection codes. A list of candidates is compiled by each team, and arranged in order of 
a numerical priority from 1 for the
best candidates to 3 or 4 for doubtful sources (e.g., ``V'' shaped transit, suspicion of secondary transits, noisy
data, mono-transits). A ``B'' is given for binary sources. All  lists are
then merged into a single list, where the sources at the top level are the candidates found by several teams at high
priorities. The teams interact regularly by means of weekly teleconferences. Apart from most likely candidates and the
binaries, all sources are rediscussed and reanalysed. 
The list of transit candidates selected by the detection teams and sorted by the probability
of their planetary nature of highest probability is then examined by the follow-up teams. 
They are responsible for confirming (or rejecting) the planetary nature of each 
candidate by ground-based observations. They focus primarily on the candidates of 
highest priorities, although stellar magnitude and amount of observing time available will 
influence their final decisions.

The transit parameters of the candidates were estimated as follows. 
First, a low-order polynomial was fit to the regions around each transit in an 
attempt to normalise the data. A first estimate of the period and epoch are used to 
phase-fold the light curve. The data points are binned, errors being
assigned according to the standard deviation of the points inside each bin divided by 
the square-root of the number of points in each bin.
A Levenberg-Marquardt algorithm \citep{Levenberg1944,Marquardt1963} is used to 
fit a trapezoid (where its center, depth, duration, and time of ingress are the 
fit parameters) to the phase-folded curve. The best-fit model trapezoid is then 
cross-correlated at each individual transit in the light curve, to determine 
their centers. A linear fit to
the resulting O-C diagram refines the estimations of the period and epoch. With this new
ephemeris, the process is iterated, until the ephemerides are
within the error bars of the previous values (typically one iteration is sufficient). 
The error in both the period and epoch are the formal errors in the linear fit.

Table~\ref{tab:param} lists the transit parameters for the 50         
planetary candidates: identifiers, coordinates, periods, and epochs
with their associated errors, the transit duration (in hours) and
depth (in \%), and  an estimate of the
stellar density inferred by the transit light curve fit as
explained in \cite{Seager2003}. 
This parameter combined with the other characteristics of the candidates 
(e.g., depth, duration, shape, out of transit modulation, stellar parameters) 
are used as input for the ranking of candidates given to the follow-up team.
We note that the light curve of the candidate 37 is contaminated with the curve of a
clear eclipsing binary (source~97 in Table~\ref{tab:binaries}). The value of its
transit parameters may therefore have been affected. 

\begin{table*}
 \centering
 \caption{Transit parameters of the 50 planetary candidates} 
  \label{tab:param}
 \begin{tabular}{rrrrrrrrrr}
 \hline
  No. & Win-ID & Period (d) & Error period (d) & Epoch (d) & Error epoch (d) &  Duration (h) & Depth & Density ($\rho_\odot$) \\
    &   &   &   &  +2454000 &   &    &  &   \\
 \hline  
 1 & E1 2046 &      -   &    -    & 167.9153 & 2.1E-03 &  5.638 & 1.2E-02 &      -   \\
 2 & E1 1319 &  1.69851 & 2.6E-05 & 136.4886 & 8.5E-04 &  2.335 & 4.4E-03 &   0.5374 \\
 3 & E1 1158 & 10.53096 & 1.3E-04 & 140.0856 & 3.3E-04 &  2.159 & 1.4E-02 &   8.9253 \\
 4 & E1 0288 &  7.89296 & 4.6E-04 & 135.0691 & 2.3E-03 &  4.512 & 3.7E-03 &   0.3740 \\
 5 & E2 3787 &  0.85809 & 3.9E-06 & 138.1467 & 1.4E-04 &  2.660 & 1.3E-03 &   0.1955 \\
 6 & E2 1857 &  0.82169 & 1.4E-05 & 138.3168 & 6.6E-04 &  1.829 & 4.7E-03 &   0.8132 \\
 7 & E1 4591 &  4.29539 & 3.5E-05 & 136.6231 & 2.6E-04 &  2.295 & 2.9E-03 &   1.0074 \\
 8 & E2 1136 &  1.22387 & 3.4E-06 & 139.2401 & 1.3E-04 &  2.748 & 1.8E-03 &   0.2076 \\
 9 & E2 2430 &  3.58747 & 9.2E-05 & 139.2341 & 8.4E-04 &  5.570 & 1.2E-02 &   0.5040 \\
10 & E2 0203 &  5.16868 & 1.9E-05 & 138.7811 & 1.2E-04 &  2.918 & 3.4E-02 &   5.3662 \\
11 & E2 1712 &  2.76741 & 5.8E-05 & 139.6140 & 5.3E-04 &  4.114 & 2.4E-03 &   0.6866 \\
12 & E1 0399 & 33.06200 & 3.5E-03 & 151.7875 & 2.2E-03 &  3.015 & 1.5E-02 &  11.0820 \\
13 & E2 1736 & 21.72025 & 1.5E-03 & 144.2915 & 2.8E-03 & 13.175 & 1.2E-02 &   0.1613 \\
14 & E1 0396 &  7.82394 & 6.9E-04 & 140.0779 & 2.7E-03 &  2.788 & 8.2E-04 &   0.6066 \\
15 & E2 1677 &      -   &    -    & 156.8022 & 1.2E-03 &  6.795 & 3.0E-02 &      -   \\
16 & E2 1126 &  1.50900 & 1.2E-05 & 138.3265 & 3.3E-04 &  2.450 & 2.2E-02 &   2.7151 \\
17 & E1 0783 &      -   &    -    & 162.9538 & 1.3E-03 &  5.498 & 6.4E-03 &      -   \\
18 & E1 0330 &  9.20191 & 3.4E-04 & 141.3652 & 1.3E-03 &  4.404 & 1.2E-02 &   3.1234 \\
19 & E1 2755 &  4.39125 & 4.2E-05 & 139.3811 & 3.9E-04 &  2.486 & 2.4E-02 &   3.8575 \\
20 & E1 4617 & 19.75581 & 3.8E-03 & 143.8531 & 3.1E-03 & 16.595 & 4.0E-02 &   0.1324 \\
21 & E2 3724 & 12.32616 & 1.4E-03 & 142.4015 & 2.7E-03 & 11.802 & 1.0E-02 &   0.1094 \\
22 & E2 4290 &  2.20546 & 1.5E-05 & 139.6775 & 2.1E-04 &  8.741 & 4.0E-03 &   0.0307 \\
23 & E1 4108 &  7.36644 & 8.4E-04 & 137.9420 & 2.7E-03 &  2.911 & 5.2E-03 &   1.6112 \\
24 & E1 1531 &  2.38147 & 6.7E-05 & 137.2002 & 8.8E-04 &  2.160 & 1.2E-02 &   2.0203 \\
25 & E2 2009 & 10.84581 & 1.4E-03 & 141.7762 & 2.9E-03 &  5.040 & 4.1E-03 &   0.2817 \\
26 & E1 2774 &  1.60551 & 5.9E-05 & 135.8954 & 1.3E-03 &  3.214 & 7.2E-03 &   0.3168 \\
27 & E2 3010 & 23.20918 & 6.1E-03 & 159.0750 & 2.6E-03 &  3.182 & 1.7E-02 &   9.3127 \\
28 & E2 4300 &  5.80656 & 3.7E-04 & 139.0526 & 1.6E-03 &  3.236 & 5.1E-03 &   0.8538 \\
29 & E2 3150 &      -   &    -    & 163.4256 & 2.1E-03 &  4.041 & 1.7E-02 &      -   \\
30 & E2 2604 &  3.81967 & 1.4E-04 & 138.2163 & 1.3E-03 &  4.366 & 2.5E-03 &   0.2051 \\
31 & E1 2648 &  3.68241 & 2.5E-04 & 138.4929 & 1.8E-03 &  3.138 & 8.8E-03 &   0.7965 \\
32 & E1 2328 &  4.50975 & 3.1E-04 & 137.6290 & 2.1E-03 &  5.911 & 6.8E-03 &   0.1238 \\
33 & E1 4998 & 10.08309 & 1.1E-03 & 142.3704 & 2.6E-03 &  2.787 & 1.9E-02 &   5.6972 \\
34 & E2 4058 &      -   &    -    & 188.9298 & 4.2E-03 &  3.651 & 9.3E-03 &      -   \\
35 & E2 3425 &  1.18553 & 3.0E-05 & 139.2383 & 9.1E-04 &  2.996 & 3.6E-03 &   0.2295 \\
36 & E2 3854 &  1.14181 & 3.0E-05 & 138.9064 & 7.4E-04 &  1.971 & 1.4E-03 &   1.4261 \\
37 & E2 3952 & 13.47756 & 4.1E-03 & 160.6192 & 3.2E-03 &  2.605 & 2.2E-03 &   7.4800 \\
38 & E2 1407 &  5.16776 & 3.2E-04 & 140.9219 & 1.8E-03 &  1.604 & 2.5E-02 &  16.2410 \\
39 & E2 2721 &  0.61161 & 6.9E-06 & 138.5715 & 3.5E-04 &  2.569 & 6.0E-03 &   0.4197 \\
40 & E2 0704 &  2.15520 & 6.1E-05 & 139.2776 & 7.9E-04 &  5.891 & 7.2E-03 &   0.1898 \\
41 & E2 4073 & 15.00128 & 1.3E-03 & 140.1756 & 2.4E-03 &  5.347 & 3.9E-02 &   2.7669 \\
42 & E1 2329 &  1.86725 & 5.6E-05 & 135.5561 & 5.8E-04 &  2.636 & 3.7E-03 &   0.4757 \\
43 & E1 3336 &  1.38972 & 3.3E-05 & 135.8757 & 7.7E-04 &  2.751 & 1.7E-03 &   0.2017 \\
44 & E1 4911 &  2.16638 & 8.7E-05 & 136.5974 & 1.3E-03 &  5.891 & 9.6E-03 &   0.1008 \\
45 & E2 4339 &  1.36204 & 3.9E-05 & 139.1842 & 9.9E-04 &  2.220 & 1.7E-03 &   0.3215 \\
46 & E2 4124 &  1.50872 & 7.0E-05 & 139.5222 & 1.3E-03 &  3.350 & 2.1E-03 &   0.1436 \\
47 & E2 3819 &  1.56554 & 4.7E-05 & 138.7047 & 8.6E-04 &  3.204 & 2.1E-02 &   0.6651 \\
48 & E2 4467 & 16.44935 & 2.2E-03 & 140.8322 & 3.0E-03 &  5.527 & 1.4E-02 &   0.9068 \\
49 & E2 3856 & 16.56276 & 1.7E-03 & 145.6439 & 2.7E-03 &  1.482 & 2.1E-02 &  70.9060 \\
50 & E2 1063 &      -   &    -    & 171.7411 & 1.5E-03 &  8.554 & 7.7E-03 &      -   \\

\hline
  \end{tabular}
\end{table*}

\begin{figure}
 \resizebox{\hsize}{!}{\includegraphics[]{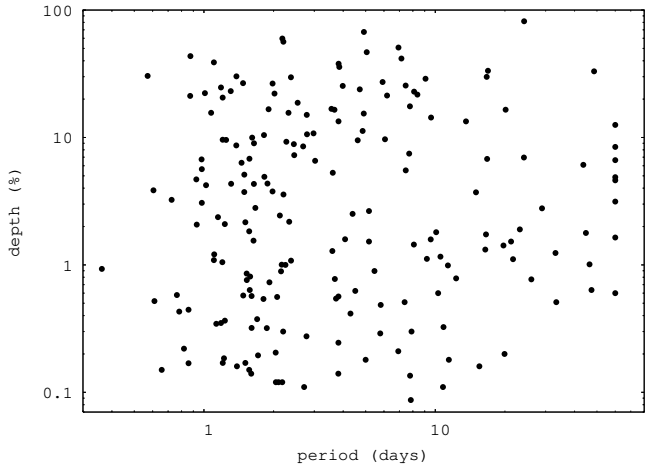}} \caption{Transit depth versus orbital period for 
 all sources with detected transits (planetary candidates and clear stellar binaries).} 
\label{fig:depth-per}
\end{figure}

\begin{figure}
 \resizebox{\hsize}{!}{\includegraphics[]{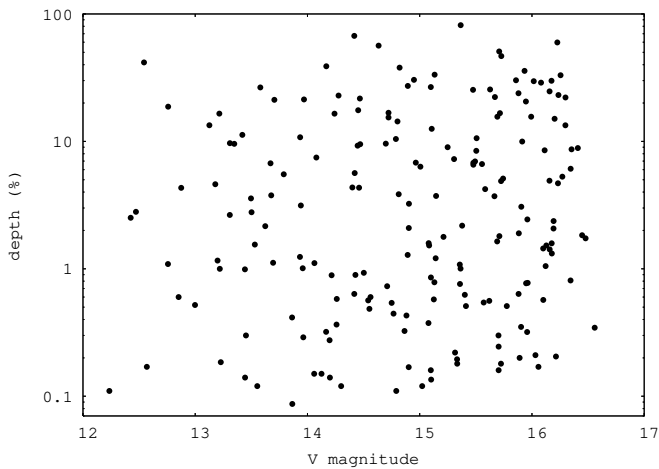}} \caption{Transit depth versus V magnitude for 
 all sources with detected transits (planetary candidates and clear stellar binaries).} 
\label{fig:depth-mag}
\end{figure}

Figure~\ref{fig:depth-per} shows the transit depth versus orbital period
for all sources in IRa01 including planetary candidates and stellar binaries. 
There does not seem to be any correlation between transit depth and
period, for periods below 10 days. The  correlation of the depth with
the  number of observed transits is evident for period $>$\,10\,days. We note that several mono-transits have
also been reported. This suggests that the detection methods used by
the detection teams do not strongly depend on                      
the number of transits as long as several are detectable. A detailed study of
the capabilities of the detection algorithms is currently ongoing.     

\begin{figure}
 \resizebox{\hsize}{!}{\includegraphics[angle=90]{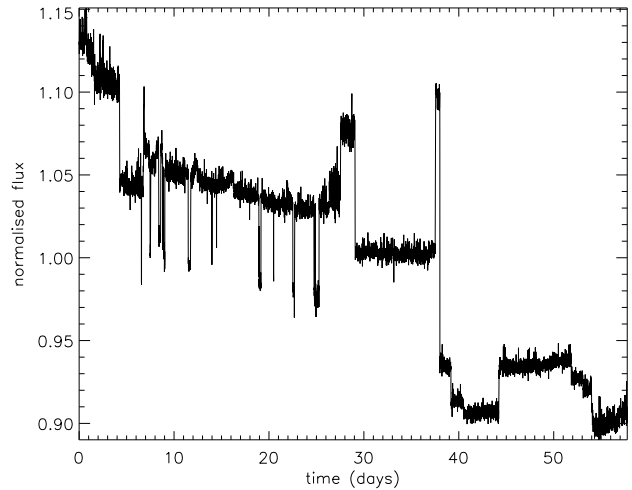}} \caption{Typical light curve 
 containing frequent jumps caused by `hot pixels'.} 
\label{fig:hotpix}
\end{figure}

Figure~\ref{fig:depth-mag} shows the same diagram but for the transit
depth versus the V magnitude. 
There is again no strong dependence between these two parameters that is
apparent for magnitudes brighter than 16, a slight dependence is however evident 
for fainter stars. This might indicate that the noise is not dominated by 
photon noise but rather by instrumental effects, including hot pixels 
(see Fig.~\ref{fig:hotpix}).

Hot pixels are characterized by sudden jumps in the light curve, followed either by an exponential or 
sudden decay. They are caused by high-energy particle impacts, mainly protons, on the detector. A description
of the radiation effects on the \corot CCD can be found in \cite{Pinheiro2008}. The number of hot pixels of intensity
higher than a certain quantity of electrons at the beginning of the first 5 \corot runs (IRa01, SRc01, LRc01, LRa01, and
SRa01) are shown by \cite{Auvergne2009} in their Figure 6. In the
case of the initial run, about 26700, 3200, and 24
bright pixels were reported with an intensity of electrons higher than 300~e$^-$, 1000~e$^-$, and 10000~e$^-$,
respectively. No efficient filtering method has so far been found that is capable of removing these sudden jumps/decays from
the light curves while leaving the transits intact. The detection teams deal
with them mainly by
renormalising the light curve before and after the jumps and leaving a gap at the place of the discontinuities. 
Replacing hot pixel events with short gaps avoids the detection of spurious signals
without having a large impact on the detected transits.
 
A  study of the noise properties was performed by \cite{Aigrain2009}. They claim that, after
pre-processing of the light curves  to minimize long-term variations and outliers, the behaviour of the
noise on a 2h timescale is close to pre-launch specification. However, a noise level of a factor 2-3 above the photon noise is
still found because of the residual jitter noise and hot pixel events. Furthermore, there is evidence of a slight
degradation in the performance over time for the first 3 long runs (IRa01, LRc01, and LRa01).

The transit detection threshold is discussed in Moutou~et~al.~2009 (accepted), following the model described 
in \cite{Pont2006}. 
In Moutou et al. 2009, they examine the location of planet candidates in the magnitude 
versus transit signal ($dn^{0.5}$, where $d$ is the transit depth and $n$ is the number of points in the transit). 
They find that the detection threshold does not depend on magnitude and conclude that
correlated fluctuations (instrumental effects or stellar variability) dominates, which is similar
to what we conclude from Fig.~\ref{fig:depth-mag}.
The detection limit is at
$dn^{0.5}=0.009$, substantially higher than in the pre-launch models.

The implications of these noise properties and detection threshold on planet detection are discussed in Fressin 
et al. 2009 (in prep.). They use the CoRoTlux transit survey simulator described in \cite{Fressin2007} to show that the CoRoT 
yield on the first 4 fields is less than one-half that expected. 
This gap will probably be reduced as the follow-up of \corot candidates nears completion.
\cite{Fressin2007}  provides an estimate of the planet occurrence 
in close orbit around F-G-K dwarf stars as a function of the radius of the planet, which agrees with radial velocity, 
ground-based transit, and \corot
discoveries. Interestingly, they show that \corot's detection of one Super-Earth (i.e., CoRoT-7b submitted by L\'eger~et~al.~2009) 
agrees with the high expectations from the HARPS team for the number of close-in Super-Earths (i.e., for 30~\% 
of main-sequence dwarfs - see \cite{Lovis2009}), because this kind of planets typically needs to have a bright K dwarf 
host to exceed the \corot detection threshold.

% ___________________________________________________________________________

\section{Summary}
\label{sec:conc}

\corot has observed its first star field, IRa01, for 2 months since the
beginning of 2008. It has obtained light curves of 3898 chromatic sources
and 5974 monochromatic sources, which have been analysed by the
detection teams. About one hundred sources have been classified as
potential candidates and 50 of them have been kept as good
candidates. The transit parameters of these candidates are listed in
Table~\ref{tab:param}. About 40 of these should be
followed-up with ground-based facilities. So far only two planets,
\corot-1b and \corot-4b, have been confirmed, from IRa01, 
each published individually as the subject of a dedicated study. 
We provide in the
Appendix a list of eclipsing binaries found in the field.

% ____________________________________________________________________________
%\begin{acknowledgement}
%\end{acknowledgement}

% ____________________________________________________________________________
\bibliographystyle{aa}
\bibliography{11882}

% ____________________________________________________________________________
\begin{appendix}
\section{Binary stars in \corot IRa01 field}
\label{sec:app}
Table~\ref{tab:binaries} lists of all eclipsing
binaries that have been identified in CoRoT-IRa01 field. Five of
these sources (No. 4, 32, 34, 97, 123) were reported in
\cite{Kabath2007} within their Berlin Exoplanet Search Telescope
(BEST) survey of variable stars in the \corot fields. Sources 1 to 139
are ordinary eclipsing binaries (note that sources labeled 39 and 40
are two binaries in the same mask of \corot, so there is a single
\corot identifier for both), whereas sources 140 to 145 are eclipsing 
binaries where only one eclipse has been found, so their period could
not be determined (these are the so-called mono-transit events).

\onecolumn

\begin{landscape}
\begin{longtable}{rrrrrrrrrr}
 \caption{Eclipsing binary candidates found in IRa01}\label{tab:binaries} \\
 \hline
   \multicolumn{1}{c}{\textbf{No.}} &
   \multicolumn{1}{c}{\textbf{CoRoT-ID}} &
   \multicolumn{1}{c}{\textbf{Win-ID}} &
   \multicolumn{1}{c}{\textbf{Alpha ($^\circ$)}} &
   \multicolumn{1}{c}{\textbf{Delta ($^\circ$)}} &
   \multicolumn{1}{c}{\textbf{V Mag}} &
   \multicolumn{1}{c}{\textbf{Period (d)}} &
   \multicolumn{1}{c}{\textbf{Epoch (d)}} &
   \multicolumn{1}{c}{\textbf{Dur. (h)}} &
   \multicolumn{1}{c}{\textbf{Depth (\%)}} \\
   \multicolumn{1}{c}{} &         % No
   \multicolumn{1}{c}{} &         % CoRoT-ID
   \multicolumn{1}{c}{} &         % Win-ID
   \multicolumn{1}{c}{} &         % Alpha
   \multicolumn{1}{c}{} &         % Delta
   \multicolumn{1}{c}{} &         % V mag
   \multicolumn{1}{c}{} &         % period
   \multicolumn{1}{c}{+2454000} & % epoch
   \multicolumn{1}{c}{} &         % dur
   \multicolumn{1}{c}{} \\        % depth
   \hline\\
\endfirsthead

%This is the header for the remaining page(s) of the table...
\multicolumn{10}{c}{{\tablename} \thetable{} -- Continued} \\
 \hline
   \multicolumn{1}{c}{\textbf{No.}} &
   \multicolumn{1}{c}{\textbf{CoRoT-ID}} &
   \multicolumn{1}{c}{\textbf{Win-ID}} &
   \multicolumn{1}{c}{\textbf{Alpha ($^\circ$)}} &
   \multicolumn{1}{c}{\textbf{Delta ($^\circ$)}} &
   \multicolumn{1}{c}{\textbf{V Mag}} &
   \multicolumn{1}{c}{\textbf{Period (d)}} &
   \multicolumn{1}{c}{\textbf{Epoch (d)}} &
   \multicolumn{1}{c}{\textbf{Dur. (h)}} &
   \multicolumn{1}{c}{\textbf{Depth (\%)}} \\
   \multicolumn{1}{c}{} &         % No
   \multicolumn{1}{c}{} &         % CoRoT-ID
   \multicolumn{1}{c}{} &         % Win-ID
   \multicolumn{1}{c}{} &         % Alpha
   \multicolumn{1}{c}{} &         % Delta
   \multicolumn{1}{c}{} &         % V mag
   \multicolumn{1}{c}{} &         % period
   \multicolumn{1}{c}{+2454000} & % epoch
   \multicolumn{1}{c}{} &         % dur
   \multicolumn{1}{c}{} \\        % depth
   \hline\\
\endhead

\\\hline
\endfoot

  1 & 102759638 &   E2 3724 &  101.248 &  -2.61254 &   15.13 &  12.32900 $\pm$  1.00E-03 &  142.39700 $\pm$ 1.59E-01 &   11.440 &    0.800 \\
  2 & 102781577 &   E2 3479 &  101.371 &  -2.93473 &   15.33 &   5.00429 $\pm$  1.61E-02 &  141.17440 $\pm$ 1.17E-01 &    7.200 &    0.300 \\
  3 & 102819749 &   E2 1568 &  101.639 &  -2.59904 &   15.11 &  36.92300 $\pm$  1.00E-03 &  162.36000 $\pm$ 1.00E-02 &    7.300 &   13.000 \\
  4 & 102941623 &   E2 2199 &  102.343 &  -2.17165 &   15.31 &   1.65060 $\pm$  3.00E-04 &  138.99400 $\pm$ 3.00E-03 &    5.900 &    0.270 \\
  5 & 102707895 &   E1 1873 &  100.951 &  -0.96872 &   13.96 &   5.78390 $\pm$  5.00E-04 &  136.85500 $\pm$ 8.00E-03 &    9.450 &    0.430 \\
  6 & 102765395 &   E2 4693 &  101.280 &  -2.31180 &   15.88 &   1.57651 $\pm$  4.72E-04 &  139.71463 $\pm$ 7.22E-03 &    1.730 &    0.630 \\
  7 & 102855348 &   E1 0081 &  101.876 &  -1.20972 &   12.23 &   5.42500 $\pm$  2.50E-03 &  136.77300 $\pm$ 1.30E-02 &    3.000 &    0.170 \\
  8 & 102823343 &   E1 4238 &  101.665 &  -1.81256 &   15.57 &   7.44690 $\pm$  1.30E-03 &  138.54600 $\pm$ 2.00E-03 &    3.620 &    6.800 \\
  9 & 102771473 &   E2 4752 &  101.315 &  -2.24456 &   16.56 &   1.13230 $\pm$  2.46E-04 &  139.37964 $\pm$ 8.25E-03 &    2.390 &    0.480 \\
 10 & 102826074 &   E1 4597 &  101.684 &  -1.32652 &   16.25 &  48.57814 $\pm$  2.87E-03 &  140.23557 $\pm$ 1.87E-03 &   10.310 &   15.620 \\
 11 & 102725806 &   E1 1875 &  101.057 &  -1.58770 &   15.42 &   0.34180 $\pm$  1.00E-04 &  136.02320 $\pm$ 2.92E-03 &    0.000 &    8.230 \\
 12 & 102739450 &   E1 2209 &  101.135 &  -0.64769 &   15.62 &   2.07265 $\pm$  1.00E-05 &  136.16830 $\pm$ 6.83E-03 &    0.560 &    2.820 \\
 13 & 102760888 &   E1 4417 &  101.255 &  -1.70619 &   13.44 &   1.90450 $\pm$  1.90E-04 &  136.09635 $\pm$ 3.20E-03 &    2.020 &    0.160 \\
 14 & 102791304 &   E2 4439 &  101.426 &  -3.11075 &   16.30 &  13.57800 $\pm$  1.00E-03 &  149.71745 $\pm$ 7.30E-04 &   25.200 &   49.000 \\
 15 & 102794063 &   E2 3276 &  101.442 &  -2.80939 &   16.35 &   0.38194 $\pm$  5.00E-05 &  138.34893 $\pm$ 1.00E-04 &    3.720 &   20.400 \\
 16 & 102794135 &   E1 0736 &  101.442 &  -1.85562 &   14.63 &   0.26387 $\pm$  5.00E-05 &  135.10625 $\pm$ 1.00E-04 &    1.710 &    5.760 \\
 17 & 102798366 &   E1 1488 &  101.476 &  -1.25857 &   15.29 &   0.39568 $\pm$  5.00E-05 &  135.47506 $\pm$ 1.00E-04 &    3.710 &    1.830 \\
 18 & 102806220 &   E2 3454 &  101.540 &  -2.47491 &   15.92 &   0.34973 $\pm$  5.00E-04 &  138.58722 $\pm$ 1.00E-04 &    3.700 &   35.000 \\
 19 & 102806409 &   E2 4376 &  101.542 &  -3.14874 &   16.09 &   0.64735 $\pm$  5.00E-04 &  138.54850 $\pm$ 1.00E-04 &    3.300 &   20.100 \\
 20 & 102808511 &   E2 0968 &  101.560 &  -2.37561 &   13.37 &   0.23966 $\pm$  5.00E-04 &  138.22310 $\pm$ 1.00E-04 &    2.600 &   31.100 \\
 21 & 102814334 &   E1 1671 &  101.601 &  -1.59656 &   14.12 &   0.65776 $\pm$  5.00E-04 &  135.76000 $\pm$ 1.00E-02 &    3.300 &    0.190 \\
 22 & 102819924 &   E2 4711 &  101.640 &  -2.92284 &   16.17 &  16.70000 $\pm$  3.00E-03 &  149.65103 $\pm$ 2.60E-03 &   26.100 &   47.800 \\
 23 & 102821683 &   E2 2608 &  101.653 &  -2.17193 &   14.75 &   1.81094 $\pm$  3.80E-04 &  140.03110 $\pm$ 8.30E-03 &   18.700 &    2.110 \\
 24 & 102822723 &   E2 2726 &  101.661 &  -2.26234 &   14.93 &  10.12150 $\pm$  3.55E-04 &  144.23984 $\pm$ 8.50E-04 &   11.200 &   30.400 \\
 25 & 102826085 &   E2 0778 &  101.684 &  -2.24207 &   13.01 &   1.02583 $\pm$  7.00E-05 &  138.88381 $\pm$ 2.07E-03 &   10.500 &    5.100 \\
 26 & 102846142 &   E1 1128 &  101.819 &  -1.45863 &   15.04 &   0.41086 $\pm$  5.00E-04 &  135.05939 $\pm$ 1.00E-04 &    4.600 &   14.500 \\
 27 & 102870524 &   E1 2045 &  101.963 &  -1.48052 &   14.13 &   1.86800 $\pm$  1.00E-03 &  136.47937 $\pm$ 1.00E-03 &    1.100 &    0.270 \\
 28 & 102888076 &   E2 3346 &  102.068 &  -2.53040 &   15.37 &  24.19210 $\pm$  2.72E-03 &  142.60754 $\pm$ 3.00E-03 &    5.900 &   34.100 \\
 29 & 102897917 &   E1 4302 &  102.120 &  -0.72015 &   16.13 &   0.44560 $\pm$  5.00E-04 &  135.27050 $\pm$ 1.00E-04 &    4.400 &   20.700 \\
 30 & 102904593 &   E1 1080 &  102.154 &  -0.60466 &   15.13 &  16.89607 $\pm$  2.94E-04 &  142.04979 $\pm$ 3.40E-04 &    5.500 &   17.900 \\
 31 & 102910432 &   E2 1277 &  102.185 &  -2.62631 &   14.87 &   0.31221 $\pm$  5.00E-04 &  138.56415 $\pm$ 1.00E-04 &    3.400 &    4.710 \\
 32 & 102924081 &   E2 0262 &  102.254 &  -1.89538 &   12.27 &   0.37350 $\pm$  1.00E-04 &  138.25600 $\pm$ 1.00E-04 &    4.000 &   24.100 \\
 33 & 102939944 &   E2 0738 &  102.334 &  -2.10704 &   14.36 &   0.87378 $\pm$  5.00E-04 &  139.04790 $\pm$ 5.00E-03 &    2.200 &    0.240 \\
 34 & 102940723 &   E2 1704 &  102.338 &  -2.11356 &   11.73 &   0.87407 $\pm$  5.00E-04 &  138.17630 $\pm$ 5.00E-03 &    3.100 &   43.600 \\
 35 & 102943300 &   E2 0915 &  102.351 &  -2.32753 &   13.21 &  20.13885 $\pm$  7.83E-03 &  153.57863 $\pm$ 5.70E-03 &   16.100 &   25.100 \\
 36 & 102961901 &   E2 2711 &  102.443 &  -2.01828 &   15.81 &   0.42090 $\pm$  1.00E-04 &  138.30993 $\pm$ 5.00E-04 &    4.300 &   30.500 \\
 37 & 102844383 &   E1 1495 &  101.808 &  -1.33584 &   15.10 &   1.52718 $\pm$  4.85E-04 &  135.45534 $\pm$ 5.00E-04 &    2.400 &    1.100 \\
 38 & 102846496 &   E2 2095 &  101.821 &  -3.11782 &   14.77 &   0.85681 $\pm$  5.00E-04 &  138.40500 $\pm$ 5.00E-04 &    1.800 &    0.430 \\
 39 & 102842572 &   E2 2746 &  101.795 &  -2.05667 &   15.91 &   4.00256 $\pm$  5.00E-04 &  138.17830 $\pm$ 5.00E-04 &    3.300 &    9.900 \\
 40 &    -      &   -       &     -    &    -      &	  -  &   2.46350 $\pm$  5.00E-04 &  139.19670 $\pm$ 5.00E-04 &    2.500 &    2.700 \\
 41 & 102745492 &   E1 4732 &  101.168 &  -1.12118 &   16.27 &   3.60467 $\pm$  7.64E-04 &  138.95489 $\pm$ 7.50E-03 &    2.472 &    5.285 \\
 42 & 102901962 &   E1 3608 &  102.141 &  -1.68750 &   16.41 &   2.44483 $\pm$  5.17E-03 &  137.09217 $\pm$ 4.55E-02 &    4.174 &    8.865 \\
 43 & 102912741 &   E2 0254 &  102.196 &  -2.89809 &   13.35 &   1.24542 $\pm$  8.90E-05 &  138.51084 $\pm$ 2.27E-03 &    2.117 &    9.565 \\
 44 & 102817472 &   E2 1390 &  101.623 &  -2.75731 &   15.13 &   1.47876 $\pm$  3.48E-04 &  138.57094 $\pm$ 7.04E-03 &    4.702 &    0.575 \\
 45 & 102846132 &   E2 3660 &  101.819 &  -2.49772 &   16.35 &   1.38144 $\pm$  5.24E-04 &  139.38388 $\pm$ 6.70E-03 &    4.487 &    8.665 \\
 46 & 102853429 &   E2 1218 &  101.865 &  -2.79528 &   13.53 &   1.63806 $\pm$  1.16E-04 &  138.50031 $\pm$ 2.21E-03 &    3.141 &    1.550 \\
 47 & 102879429 &   E1 1827 &  102.018 &  -1.57061 &   15.48 &   4.03054 $\pm$  6.90E-05 &  137.84405 $\pm$ 1.12E-03 &    2.872 &    6.560 \\
 48 & 102982347 &   E2 1745 &  102.563 &  -2.97465 &   13.94 &   2.97762 $\pm$  4.30E-05 &  139.86651 $\pm$ 4.16E-04 &    4.113 &   10.770 \\
 49 & 102756466 &   E1 3987 &  101.230 &  -0.86304 &   15.48 &  16.76987 $\pm$  2.93E-01 &  151.94809 $\pm$ 1.26E-01 &    6.752 &    6.785 \\
 50 & 102779171 &   E1 1499 &  101.357 &  -1.06390 &   13.44 &  11.33939 $\pm$  3.72E-02 &  143.75596 $\pm$ 1.34E-01 &    7.545 &    2.000 \\
 51 & 102738614 &   E1 0827 &  101.130 &  -1.17246 &   14.45 &   7.76844 $\pm$  5.16E-02 &  135.81384 $\pm$ 1.46E-01 &    3.511 &   17.555 \\
 52 & 102818537 &   E2 2620 &  101.631 &  -2.85605 &   14.45 &   2.27296 $\pm$  4.80E-05 &  139.72413 $\pm$ 6.41E-04 &    5.438 &    9.254 \\
 53 & 102820928 &   E2 4500 &  101.648 &  -2.61561 &   16.16 &   1.82502 $\pm$  1.77E-04 &  138.45024 $\pm$ 3.15E-03 &    3.848 &    4.915 \\
 54 & 102867757 &   E2 4431 &  101.947 &  -2.68001 &   16.11 &   2.68580 $\pm$  2.89E-04 &  140.21954 $\pm$ 3.46E-03 &    2.552 &    8.510 \\
 55 & 102708916 &   E1 0484 &  100.957 &  -0.79757 &   13.97 &   6.18906 $\pm$  6.35E-02 &  141.10739 $\pm$ 3.44E-01 &    5.335 &   21.340 \\
 56 & 102726405 &   E1 0801 &  101.061 &  -1.37603 &   12.76 &   2.54204 $\pm$  2.90E-05 &  138.05727 $\pm$ 3.47E-04 &    4.280 &   18.720 \\
 57 & 102732394 &   E1 1251 &  101.095 &  -1.42804 &   14.96 &   1.57360 $\pm$  4.30E-05 &  135.74521 $\pm$ 9.19E-04 &    3.056 &    6.810 \\
 58 & 102733170 &   E1 1543 &  101.100 &  -1.61003 &   13.68 &   1.97940 $\pm$  4.50E-05 &  135.86023 $\pm$ 7.68E-04 &    3.949 &    3.770 \\
 59 & 102734453 &   E1 2507 &  101.107 &  -0.67484 &   14.47 &   8.37056 $\pm$  6.80E-05 &  144.18882 $\pm$ 3.35E-03 &   21.700 &    8.371 \\
 60 & 102735868 &   E1 3810 &  101.115 &  -1.33074 &   15.25 &   1.64709 $\pm$  1.80E-05 &  135.56796 $\pm$ 2.15E-03 &    3.775 &    9.000 \\
 61 & 102741994 &   E1 2336 &  101.149 &  -1.64399 &   14.47 &   4.62211 $\pm$  1.07E-04 &  135.78605 $\pm$ 7.22E-04 &    4.124 &    9.500 \\
 62 & 102752408 &   E1 3080 &  101.207 &  -1.12243 &   16.13 &  21.26057 $\pm$  5.41E-03 &  146.95066 $\pm$ 6.96E-03 &    4.600 &    1.525 \\
 63 & 102754263 &   E1 3846 &  101.217 &  -1.05010 &   15.31 &   2.45716 $\pm$  1.50E-04 &  136.94466 $\pm$ 1.78E-03 &    2.994 &    7.260 \\
 64 & 102756903 &   E1 4392 &  101.232 &  -1.44495 &   15.90 &   0.97909 $\pm$  1.50E-05 &  136.17055 $\pm$ 4.74E-04 &    2.418 &    3.070 \\
 65 & 102757626 &   E1 0791 &  101.236 &  -1.23750 &   14.70 &   1.20543 $\pm$  6.10E-05 &  135.91488 $\pm$ 1.60E-03 &    3.441 &    9.600 \\
 66 & 102764398 &   E2 3602 &  101.275 &  -2.81121 &   16.23 &   0.92744 $\pm$  5.00E-04 &  138.08047 $\pm$ 1.00E-04 &    2.702 &    4.689 \\
 67 & 102768859 &   E2 4148 &  101.300 &  -2.81762 &   16.10 &   8.06342 $\pm$  5.26E-04 &  141.27802 $\pm$ 5.80E-03 &    4.913 &    1.444 \\
 68 & 102773399 &   E1 2875 &  101.326 &  -0.95011 &   14.81 &   0.60560 $\pm$  1.00E-04 &  135.51522 $\pm$ 7.15E-03 &    2.702 &    3.850 \\
 69 & 102774523 &   E1 1052 &  101.332 &  -1.85324 &   14.89 &   5.91776 $\pm$  1.09E-04 &  135.67799 $\pm$ 5.51E-04 &    3.901 &   27.240 \\
 70 & 102776173 &   E2 1176 &  101.341 &  -3.21882 &   14.90 &   0.72508 $\pm$  5.00E-05 &  138.59908 $\pm$ 5.00E-04 &    2.702 &    3.240 \\
 71 & 102776386 &   E2 1137 &  101.342 &  -2.86191 &   13.50 &   2.20677 $\pm$  2.70E-04 &  139.42573 $\pm$ 3.74E-03 &    3.684 &    3.570 \\
 72 & 102776565 &   E2 2143 &  101.343 &  -3.15951 &   14.63 &   2.20584 $\pm$  6.50E-05 &  140.76363 $\pm$ 8.51E-04 &    4.111 &   56.400 \\
 73 & 102776605 &   E1 3357 &  101.344 &  -0.65412 &   16.23 &   2.18327 $\pm$  5.20E-05 &  137.38286 $\pm$ 7.53E-04 &    4.109 &   59.790 \\
 74 & 102783117 &   E1 1002 &  101.379 &  -0.66941 &   14.88 &   0.78215 $\pm$  1.00E-04 &  135.29534 $\pm$ 1.16E-03 &    2.702 &    0.430 \\
 75 & 102785724 &   E1 2613 &  101.394 &  -1.57614 &   15.88 &   4.71634 $\pm$  6.50E-05 &  139.61223 $\pm$ 4.28E-04 &    5.267 &   23.860 \\
 76 & 102790392 &   E2 1005 &  101.421 &  -2.46922 &   14.72 &   4.91014 $\pm$  9.90E-05 &  139.89439 $\pm$ 5.78E-04 &    5.419 &   15.410 \\
 77 & 102793963 &   E1 3124 &  101.441 &  -1.62531 &   16.19 &   1.24225 $\pm$  1.00E-04 &  135.40128 $\pm$ 5.95E-04 &    1.596 &    2.070 \\
 78 & 102802054 &   E2 4445 &  101.506 &  -2.33150 &   15.96 &   2.12884 $\pm$  1.69E-04 &  140.44080 $\pm$ 2.20E-03 &    2.540 &    2.440 \\
 79 & 102803023 &   E1 4206 &  101.514 &  -0.64875 &   15.69 &   2.32061 $\pm$  6.00E-05 &  137.82606 $\pm$ 8.25E-04 &    4.263 &   15.620 \\
 80 & 102806377 &   E2 0836 &  101.541 &  -2.03210 &   13.13 &   3.81666 $\pm$  3.70E-03 &  142.04834 $\pm$ 2.16E-02 &    4.222 &   13.380 \\
 81 & 102806577 &   E2 1918 &  101.543 &  -3.23352 &   14.24 &   3.66704 $\pm$  3.20E-04 &  140.43028 $\pm$ 2.57E-03 &    4.925 &   16.500 \\
 82 & 102809393 &   E2 0486 &  101.566 &  -2.83432 &   14.08 &   7.71063 $\pm$  4.54E-02 &  139.08695 $\pm$ 5.58E-02 &    6.247 &    7.470 \\
 83 & 102811578 &   E2 0416 &  101.582 &  -1.98315 &   12.47 &   1.66868 $\pm$  1.25E-04 &  139.03337 $\pm$ 2.25E-03 &    2.782 &    2.800 \\
 84 & 102813089 &   E1 4561 &  101.592 &  -0.99226 &   16.24 &   1.30626 $\pm$  3.90E-05 &  136.71588 $\pm$ 9.15E-04 &    3.026 &   23.100 \\
 85 & 102816070 &   E2 2295 &  101.613 &  -2.31198 &   15.63 &   7.44703 $\pm$  2.65E-04 &  140.40955 $\pm$ 8.91E-04 &    6.238 &   25.550 \\
 86 & 102818428 &   E2 1307 &  101.630 &  -2.24680 &   13.79 &   7.45491 $\pm$  4.48E-04 &  142.37746 $\pm$ 1.54E-03 &    6.238 &    5.520 \\
 87 & 102819360 &   E2 3054 &  101.636 &  -2.86079 &   15.15 &   0.99596 $\pm$  5.00E-04 &  138.59319 $\pm$ 5.60E-03 &    2.812 &    3.730 \\
 88 & 102819692 &   E1 3127 &  101.638 &  -1.59328 &   15.86 &   1.38268 $\pm$  2.10E-05 &  136.22874 $\pm$ 4.95E-04 &    3.462 &   30.170 \\
 89 & 102824749 &   E1 1971 &  101.675 &  -0.75888 &   14.28 &   8.09754 $\pm$  2.55E-04 &  139.29042 $\pm$ 9.48E-04 &    6.403 &   22.900 \\
 90 & 102826984 &   E1 3686 &  101.691 &  -0.68487 &   15.10 &   1.47677 $\pm$  1.40E-05 &  136.83385 $\pm$ 3.55E-04 &    3.046 &   26.680 \\
 91 & 102828417 &   E2 1036 &  101.701 &  -2.04626 &   14.80 &   9.59460 $\pm$  2.66E-04 &  147.32365 $\pm$ 7.26E-04 &    6.453 &   14.340 \\
 92 & 102835452 &   E2 4071 &  101.748 &  -2.22388 &   15.71 &   6.93290 $\pm$  7.68E-03 &  139.00417 $\pm$ 1.69E-02 &    5.080 &   50.780 \\
 93 & 102836138 &   E1 0844 &  101.753 &  -1.34797 &   14.72 &   3.55818 $\pm$  4.90E-05 &  137.20937 $\pm$ 3.67E-04 &    3.922 &   16.770 \\
 94 & 102836169 &   E2 4009 &  101.753 &  -2.56595 &   16.16 &   1.18554 $\pm$  5.00E-05 &  139.23407 $\pm$ 1.40E-03 &    1.874 &   24.690 \\
 95 & 102840080 &   E2 3619 &  101.779 &  -2.93765 &   15.38 &   2.33737 $\pm$  1.93E-04 &  139.61448 $\pm$ 2.40E-03 &    2.557 &    2.180 \\
 96 & 102841939 &   E1 5038 &  101.791 &  -0.55531 &   16.02 &   2.37762 $\pm$  3.60E-05 &  135.82534 $\pm$ 4.82E-04 &    3.983 &   29.670 \\
 97 & 102842120 &   E2 3952 &  101.792 &  -2.95054 &   14.17 &   1.10449 $\pm$  3.90E-05 &  138.62653 $\pm$ 1.13E-03 &    3.285 &   38.860 \\
 98 & 102842466 &   E1 3571 &  101.795 &  -1.03743 &   14.42 &   4.91740 $\pm$  5.04E-04 &  138.74857 $\pm$ 2.86E-03 &    5.419 &   67.290 \\
 99 & 102844991 &   E1 3252 &  101.812 &  -0.91820 &   15.99 &   1.07425 $\pm$  1.80E-05 &  135.53889 $\pm$ 5.45E-04 &    2.428 &   15.620 \\
100 & 102849348 &   E2 2452 &  101.840 &  -2.82048 &   14.79 &   1.81837 $\pm$  5.60E-05 &  138.60443 $\pm$ 9.44E-04 &    3.223 &   10.440 \\
101 & 102851363 &   E2 3081 &  101.852 &  -2.25781 &   15.67 &   1.01078 $\pm$  2.10E-05 &  139.00262 $\pm$ 6.33E-04 &    2.419 &   22.270 \\
102 & 102852229 &   E2 0872 &  101.858 &  -2.77432 &   13.31 &   6.06061 $\pm$  3.82E-02 &  140.73996 $\pm$ 1.83E-01 &    5.756 &    9.680 \\
103 & 102858100 &   E2 2099 &  101.892 &  -3.07796 &   14.46 &   1.31134 $\pm$  9.00E-05 &  138.64401 $\pm$ 2.12E-03 &    3.311 &    4.330 \\
104 & 102870155 &   E2 4907 &  101.961 &  -2.81992 &   16.20 &   2.78294 $\pm$  8.40E-05 &  139.98194 $\pm$ 8.08E-04 &    4.440 &   15.050 \\
105 & 102870613 &   E2 0117 &  101.964 &  -2.68863 &   12.54 &   7.13947 $\pm$  4.85E-04 &  145.98116 $\pm$ 1.75E-03 &    6.084 &   41.650 \\
106 & 102870852 &   E2 0609 &  101.965 &  -2.90340 &   14.26 &   0.76471 $\pm$  5.20E-04 &  138.56904 $\pm$ 1.89E-03 &    2.844 &    2.500 \\
107 & 102872646 &   E2 0818 &  101.976 &  -2.72664 &   14.40 &   1.88286 $\pm$  4.17E-04 &  138.92797 $\pm$ 6.80E-05 &    3.940 &    4.350 \\
108 & 102879375 &   E2 0365 &  102.017 &  -2.09075 &   13.67 &   0.97728 $\pm$  5.00E-04 &  138.77715 $\pm$ 3.43E-03 &    3.129 &    6.730 \\
109 & 102882044 &   E1 3079 &  102.033 &  -0.48467 &   16.08 &   9.07345 $\pm$  5.60E-05 &  136.91781 $\pm$ 1.80E-05 &    4.127 &   28.910 \\
110 & 102884662 &   E1 1938 &  102.048 &  -1.00089 &   15.93 &   3.84822 $\pm$  2.16E-04 &  138.97463 $\pm$ 1.73E-03 &    4.935 &   35.770 \\
111 & 102886012 &   E1 4690 &  102.056 &  -1.61420 &   16.34 &   1.58466 $\pm$  2.52E-04 &  136.15763 $\pm$ 4.04E-03 &    2.631 &    0.810 \\
112 & 102889458 &   E1 4646 &  102.075 &  -1.00521 &   16.30 &   2.01989 $\pm$  3.30E-05 &  136.31380 $\pm$ 5.05E-04 &    3.953 &   22.100 \\
113 & 102892869 &   E1 3024 &  102.093 &  -0.56110 &   15.08 &   4.07590 $\pm$  6.59E-04 &  139.53177 $\pm$ 5.41E-03 &    3.810 &    1.589 \\
114 & 102896719 &   E1 3444 &  102.114 &  -0.47842 &   14.90 &   1.23114 $\pm$  1.10E-04 &  135.97699 $\pm$ 2.98E-03 &    3.444 &    2.090 \\
115 & 102900859 &   E1 1220 &  102.135 &  -0.53002 &   13.42 &   4.85346 $\pm$  1.48E-04 &  136.52903 $\pm$ 8.29E-04 &    5.416 &   11.250 \\
116 & 102902696 &   E1 1276 &  102.145 &  -0.75803 &   13.58 &   1.98087 $\pm$  1.40E-05 &  136.66216 $\pm$ 2.41E-04 &    3.665 &   26.510 \\
117 & 102914654 &   E2 4083 &  102.206 &  -2.15859 &   15.95 &   1.20531 $\pm$  1.80E-05 &  139.12553 $\pm$ 4.68E-04 &    3.441 &   20.550 \\
118 & 102929837 &   E2 1064 &  102.283 &  -2.89422 &   14.82 &   3.81996 $\pm$  4.10E-05 &  139.41564 $\pm$ 3.47E-04 &    4.934 &   37.900 \\
119 & 102930316 &   E2 2382 &  102.286 &  -2.95754 &   15.74 &   1.49577 $\pm$  1.26E-04 &  139.76033 $\pm$ 2.53E-03 &    3.617 &    5.110 \\
120 & 102931335 &   E1 3946 &  102.291 &  -1.44787 &   15.47 &   3.97923 $\pm$  2.90E-05 &  138.84505 $\pm$ 2.29E-04 &    4.516 &   25.390 \\
121 & 102932176 &   E2 1219 &  102.295 &  -1.99175 &   13.71 &   0.87225 $\pm$  1.30E-05 &  139.05420 $\pm$ 2.29E-04 &    2.987 &   21.190 \\
122 & 102943073 &   E2 0151 &  102.349 &  -2.48166 &   12.88 &   1.64410 $\pm$  7.50E-05 &  139.62393 $\pm$ 1.34E-03 &    2.921 &    4.320 \\
123 & 102955089 &   E2 1261 &  102.408 &  -1.83831 &   14.95 &   0.57161 $\pm$  1.70E-05 &  138.98060 $\pm$ 3.17E-04 &    2.702 &   30.420 \\
124 & 102961237 &   E2 3896 &  102.439 &  -2.56702 &   15.58 &   1.02102 $\pm$  6.40E-05 &  139.41378 $\pm$ 1.77E-03 &    3.132 &    4.220 \\
125 & 102965963 &   E2 4756 &  102.467 &  -2.16127 &   15.71 &   1.90365 $\pm$  3.30E-05 &  140.09132 $\pm$ 4.99E-04 &    3.942 &   16.660 \\
126 & 102980178 &   E2 4236 &  102.550 &  -2.68934 &   15.73 &   5.05476 $\pm$  1.04E-03 &  139.07148 $\pm$ 4.98E-03 &    5.426 &   46.710 \\
127 & 102983538 &   E2 2825 &  102.570 &  -2.93102 &   15.01 &   1.45435 $\pm$  1.20E-04 &  139.03886 $\pm$ 2.39E-03 &    3.612 &    6.330 \\
128 & 102801922 &   E1 0617 &  101.505 &  -1.50686 &   14.43 &   5.45907 $\pm$  4.20E-05 &  138.93015 $\pm$ 2.32E-03 &    3.503 &    0.895 \\
129 & 102927840 &   E2 4136 &  102.273 &  -2.66893 &   14.56 &  10.29535 $\pm$  8.58E-03 &  140.69140 $\pm$ 1.81E-02 &    7.055 &    0.600 \\
130 & 102726103 &   E1 0830 &  101.060 &  -1.22901 &   14.54 &   3.81466 $\pm$  2.20E-05 &  137.99633 $\pm$ 1.82E-03 &    1.894 &    0.565 \\
131 & 102786821 &   E1 2938 &  101.400 &  -1.32339 &   14.71 &   1.91876 $\pm$  1.88E-04 &  135.51978 $\pm$ 3.41E-03 &    2.969 &    0.730 \\
132 & 102802298 &   E2 4626 &  101.508 &  -2.45497 &   15.51 &   3.71918 $\pm$  5.20E-05 &  139.05857 $\pm$ 4.08E-04 &    3.578 &   10.590 \\
133 & 102932955 &   E2 2046 &  102.299 &  -2.76750 &   15.49 &  24.25906 $\pm$  4.75E-04 &  140.53257 $\pm$ 5.33E-04 &    3.960 &    6.960 \\
134 & 102806484 &   E2 4826 &  101.542 &  -2.71639 &   16.18 &   9.56081 $\pm$  4.90E-05 &  140.53819 $\pm$ 1.38E-03 &    2.729 &    1.585 \\
135 & 102735257 &   E1 3236 &  101.111 &  -1.635940 &   16.12 &   23.6935$\pm$  5.00e-04 &   156.952  $\pm$ 5.00e-04 &    4.265 &     11.6 \\
136 & 102937382 &   E2 4533 &  102.322 &  -1.909260 &	15.02 &  7.69724 $\pm$  3.89E-03 &  139.15290 $\pm$ 1.38E-03 &    6.247 &    0.170 \\
137 & 102734591 &   E1 0663 &  101.107 &  -1.291500 &	14.39 &  8.16270 $\pm$  6.13E-03 &  139.34211 $\pm$ 2.35E-02 &    6.400 &    0.100 \\
138 & 102751150 &   E2 0193 &  101.200 &  -2.087960 &	13.07 &  6.34015 $\pm$  1.50E-03 &  144.05301 $\pm$ 6.23E-03 &    5.910 &    0.160 \\
139 & 102805003 &   E2 2539 &  101.530 &  -1.548410 &	14.54 &  5.57236 $\pm$  6.45E-03 &  136.21190 $\pm$ 2.65E-03 &   13.800 &    0.110 \\
140 & 102765275 &   E1 2060 &  101.280 &  -0.671730 &   15.55 &      -                   &  177.43000 $\pm$ 1.00E-02 &    7.200 &    6.200 \\
141 & 102829388 &   E2 3914 &  101.708 &  -3.030360 &   15.50 &      -                   &  163.86000 $\pm$ 1.00E-02 &   18.400 &    8.500 \\
142 & 102855409 &   E2 1633 &  101.877 &  -2.299030 &   13.18 &      -                   &  155.43000 $\pm$ 1.00E-02 &   20.260 &   17.200 \\
143 & 102868004 &   E2 2416 &  101.949 &  -2.206550 &   15.72 &      -                   &  172.54000 $\pm$ 1.00E-02 &    5.750 &    5.700 \\
144 & 102919036 &   E1 4818 &  102.229 &  -1.072490 &   15.69 &      -                   &  174.45000 $\pm$ 1.00E-02 &   17.800 &    1.800 \\
145 & 102801672 &   E2 4912 &  101.503 &  -2.599640 &   16.34 &      -                   &  138.18000 $\pm$ 1.00E-02 &    9.070 &    8.200 \\

\end{longtable}
\end{landscape}
\end{appendix}
\end{document}